# Breaking the picomolar barrier in lateral flow assays using Bright-Dtech™ 614 - Europium nanoparticles for enhanced sensitivity

Juliette Lajoux[a,δ], Yulieth D. Banguera-Ordoñez[b,c,δ], Amadeo Sena-Torralba[b]*, Loïc J. Charbonnière,[d] Mohamadou Sy[a], Joan Goetz[a], Ángel Maquieira[b,e], Sergi Morais [b,e]∗

[a] Poly-D-Tech, 204 Avenue de Colmar, 67100 Strasbourg, France

[b] Instituto Interuniversitario de Investigación de Reconocimiento Molecular y Desarrollo Tecnológico (IDM), Universitat Politècnica de València, Universitat de València, Camino de Vera s/n, 46022, Valencia, Spain.

[c] Grupo de Investigación y Desarrollo en Ciencias, Tecnología e Innovación (BioGRID), Sociedad de Doctores e Investigadores de Colombia (SoPhIC), Bogotá, Colombia.

[d] Pluridisciplinary Hubert Curien Institute (IPHC, UMR 7178, CNRS/Université de Strasbourg), 25 rue Becquerel, 67087, Strasbourg, France.

[e] Departamento de Química, Universitat Politècnica de València, Camino de Vera s/n, 46022, Valencia, Spain.

**\*Correspondence: asentor@upvnet.upv.es (A. Sena-Torralba), smorais@upv.es (S. Morais)

[δ]The authors contributed equally to this work.

**Abstract**

Lateral flow immunoassays (LFIA) are among the most widely used rapid diagnostic tests for point-of-care screening of disease biomarkers. However, their limited sensitivity hinders their use in complex clinical applications that require accurate biomarker quantification for precise medicine. To address this limitation, we evaluated Bright-Dtech™-614 Europium nanoparticles to enhance LFIA assay sensitivity. These nanoparticles exhibited a luminescence quantum yield of 70% and a 90% conjugation efficacy with antibodies by direct adsorption. Considering these properties, we developed an LFIA to quantify human lactate dehydrogenase (h-LDH), a biomarker and therapeutic target in cancer disease. The Bright-Dtech$^{TM}$-614 Eu nanoparticle-based assay achieved a detection limit of 38 pg mL$^{-1}$, representing a 686-fold, 15-fold, and 2.9-fold improvement in sensitivity over conventional LFIA platforms using gold (AuNPs), carbon nanoparticles, and standard ELISA, respectively. The assay exhibited strong accuracy, with a mean recovery rate of 108 ± 11%, and demonstrated excellent reproducibility, as evidenced by inter- and intra-batch RSD values of 4.9% and 9.7%, respectively, when testing LDH-spiked serum samples. By substituting traditional gold nanoparticles with the Bright-Dtech™-614 Eu nanoparticles, we achieved detection limits in the femtomolar range, significantly broadening the applicability of LFIA for precision medicine.

**1. Introduction**

Lateral Flow Immunoassays (LFIA) are cost-effective paper-based assays designed to deliver rapid results at the point of sample collection [1]. Their simplicity allows use by end-users with no specialized training or additional equipment requirements, making LFIA a popular choice for home-based tests across various physiological assessments [2], including pregnancy, ovulation, and COVID-19 detection, based on visual interpretation of colorimetric signals in the test (TL) and control lines (CL) [3,4]. However, conventional LFIAs using colorimetric gold nanoparticles (AuNPs) lack the sensitivity needed for more complex clinical applications, where target biomarkers are present at ultra-low concentrations and often masked by higher-abundance plasma proteins [5]. In such cases, enhanced assay sensitivity is essential to accurately quantify biomarker levels, facilitating early diagnosis and timely therapeutic intervention [6].

Over recent decades, various strategies have been explored to improve the sensitivity of LFIA, including signal amplification and sample preconcentration techniques. However, these approaches often require multiple steps or additional equipment, limiting their practicality for point-of-care use due to increased assay complexity and cost [7,8]. A more streamlined solution is to shift from colorimetric to time-resolved fluorescence detection, which improves signal intensity at both the test and control lines by offering a higher signal-to-noise ratio [9,10]. This approach is typically achieved with lanthanide-doped nanoparticles, such as those containing Europium, Terbium, or Dysprosium, which provide superior analytical sensitivity [11,12]. Although LFIAs incorporating lanthanide nanoparticles have shown remarkable sensitivity [13–15], their use demands specialized expertise in inorganic chemistry and nanotechnology for effective synthesis, characterization, and bioreceptor conjugation, limiting their accessibility to a broader range of researchers.

To expand the accessibility of enhanced-sensitivity LFIA technology, we developed an LFIA utilizing commercially available Europium nanoparticles (Bright-Dtech™-614

Eu) and a bioconjugation kit (Link-Dtech™ 614–Eu) [16]. This approach simplifies the integration of lanthanide technology into LFIA without requiring specialized expertise. The assay detected human lactate dehydrogenase (h-LDH), a biomarker in intravascular hemolysis and vaso-occlusive crises [17-18]. Moreover, h-LDH is a significant biomarker in oncology, with elevated serum levels commonly observed in patients with a range of malignancies. These elevated levels are associated with poor clinical prognosis and resistance to therapeutic interventions. Consequently, the quantification of h-LDH has become an integral tool in diagnosing cancer and monitoring therapeutic responses [19-21].

Building on our prior work, where we developed high-affinity homemade polyclonal antibodies for h-LDH (Kd 5 nM) and demonstrated a 55-fold sensitivity increase with carbon nanoparticles (CNPs) over AuNPs [22], this study further improves sensitivity. Using Bright-Dtech™-614 Eu nanoparticles, we achieved h-LDH detection limits in the femtomolar range, which are 686-, 15-, and 2.9-fold lower than those of conventional AuNPs-based LFIA, CNPs-based LFIA, and well-plate ELISA, respectively (Scheme 1). This is the first time Bright-Dtech™-614 Europium nanoparticles are used in LFIA to boost the analytical sensitivity of the assay, achieving the lowest LoD ever reported for h-LDH, even lower than the gold standard ELISA method. The assay demonstrated a recovery rate of 108 % with high reproducibility, evidenced by inter- and intra-batch RSD values of 4.9% and 9.7% in h-LDH–spiked serum samples. These findings reveal the potential for LFIA developers to adopt this technology for significantly enhanced assay performance.

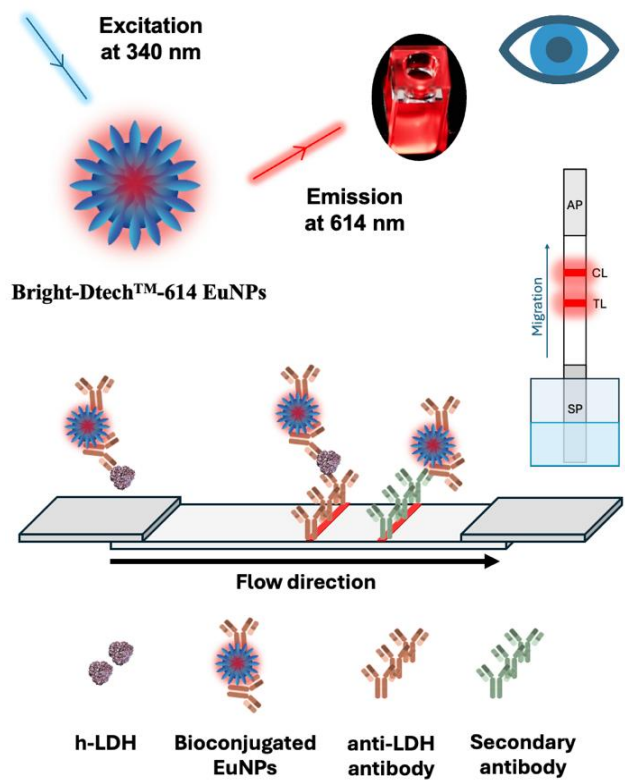

**Scheme 1.** Illustration of the Bright-Dtech™ 614 - Eu (EuNPs) that emit a red fluorescent light at 614 nm when excited at 340 nm. Depiction of the LFIA assay for detecting h-LDH using a non-competitive assay format with the Bright-Dtech™-614 Eu NPs as signal reporters.

## 2. Materials and methods

### 2.1. *Reagents, materials and instruments*

The reagents, materials, and instruments are described in the supplementary material.

### *2.2. Bright-Dtech™ 614 - Eu (EuNPs) characterization*

Transmission electron microscopy (TEM) images were acquired on a JEOL 2100F electron microscope at 200 kV, equipped with a GATAN GIF 200 electron imaging filter. Powder samples were dispersed in water, and a drop of the suspension was deposited onto TEM grids prepared with a porous membrane coated with an amorphous carbon layer. Only Bright-Dtech™-614 Eu nanoparticles located on the strand edges of membrane holes were analyzed to minimize background noise from the amorphous carbon. Images were processed using ImageJ software, and core diameters were measured on approximately 100 nanoparticles. The straight-line tool was used to measure nanoparticle diameters from edge to edge. Luminescence quantum yield was quantified using a reference fluorophore with known quantum yield and the formula reported by Valeur, B[16].

Excitation and emission spectra were recorded using a TECAN Spark microplate reader equipped with a high-energy Xenon flash lamp and EuNPs at a concentration of 1 nM. Emission spectra were collected under time-resolved fluorescence conditions with a 70 µs delay time, an excitation wavelength of 340 nm, and a 1000 µs integration time. Similarly, excitation spectra were obtained with a 70 µs delay time, an emission wavelength of 610 nm, and a 1000 µs integration time.

### *2.3. EuNPs bioconjugation with anti-LDH antibody*

The Link-Dtech™ 614–Eu coupling protocol was used to conjugate Bright-Dtech™ 614 Eu nanoparticles (EuNPs) with anti-LDH antibodies. First, EuNPs were resuspended to a concentration of $1 \times 10^{-8}$ M in coupling buffer, then homogenized by sonicating for 10 s at room temperature or by vigorous vortexing. The anti-LDH antibodies were then added at a nanoparticle-to-antibody ratio of 150:1, and the mixture was incubated overnight at 4 ºC. After incubation, the solution was centrifuged, the supernatant was removed, and the pellet was resuspended in the coupling buffer. The final conjugate solution was diluted to a $2 \times 10^{-8}$ M in the coupling buffer. Conjugation efficacy was assessed by comparing the concentration of antibodies in the supernatant before and after centrifugation.

### *2.4. Fabrication of LFIA strips*

The anti-LDH antibody (0.8 mg/mL in PBS with 0.1% BSA) and goat anti-rabbit antibody (0.5 mg/mL in PBS) were dispensed onto the nitrocellulose (NC) membrane at the TL and CL, respectively, using an Agismart Rapid Test Printer (Regabiotech). Following overnight incubation at room temperature, the NC membrane was assembled with a sample pad and absorbent pad. The laminated membrane was then cut into 4 mm-wide strips.

### 2.5. h-LDH detection and analytical validation

Calibration curves for h-LDH detection were generated by adding 75 µL of serially diluted h-LDH (0 to 100 ng/mL in migration buffer) and 5.0 µL of Bright-Dtech™ 614 EuNPs (NPs diluted at $7.6 \times 10^{-10}$ M in migration buffer) into wells of a 96-well plate. LFIA strips were then immersed in each well for sample migration. After 20 minutes, images of the strips were captured using the Molecular Devices SpectraMax ID5 plate reader with its western blot module under constant lighting. Signal intensities at the TL,

CL, and background (BG) were quantified with ImageJ software [1]. Normalization was performed using the TL/CL ratio, and the data were fitted to a four-parameter logistic (sigmoidal) model.

Analytical parameters, including the limit of detection (LoD), limit of quantification (LoQ), EC50, linear dynamic range (LDR), and linear regression coefficient (r²), were determined. The LoD was calculated as the optical intensity of the blank sample plus three times its standard deviation (LoD = blank + 3σ_blank). In comparison, the LoQ was calculated as the blank intensity plus ten times its standard deviation (LoQ = blank + 10σ_blank) [23]. The LDR was defined as the range corresponding to a signal intensity between 10% and 90% of the maximum signal output [24]. Recovery studies were conducted on goat serum samples spiked with known concentrations of h-LDH. The percentage recovery was calculated by interpolating the signals on the calibration curve and comparing the detected concentration with the expected one.

## 3. Results and discussion

### 3.1. Characterization of the EuNPs

Bright-Dtech™ 614 EuNPs are commercially available lanthanide-based nanoparticles containing europium in their core. They are designed to emit time-resolved fluorescence, enhancing the signal-to-noise ratio at the test line in LFIA applications. Before their use in LFIA, EuNPs were characterized to confirm that their size and optical properties were suitable for assay integration. TEM analysis showed that the EuNPs were spherical with diameters under 100 nm (Figure 1A), a size critical for efficient flow through the strip's detection pad (pore size ~10 μm). A histogram of the size distribution revealed a homogeneous dispersion with an average diameter of 53 ± 12 nm (Figure 1B). The stability of Bright-Dtech™ 614 Eu nanoparticles was evaluated by monitoring their hydrodynamic size using Dynamic Light Scattering (DLS) over two years under storage conditions of 23 ºC and 4 ºC. The results in Figure S1 demonstrate no significant variations in nanoparticle size between the two storage temperatures over the two-year duration. These findings indicate that the nanoparticles exhibit long-term stability under the tested conditions.

Optical characterization of EuNPs indicated two excitation peaks at 340 and 280 nm. Upon excitation at 340 nm, three sharp emission peaks were observed at 590, 614, and 690 nm, with the highest fluorescence intensity at 590 nm (Figure 1C). These properties confirm the suitability of EuNPs as a high-sensitivity signal reporter.

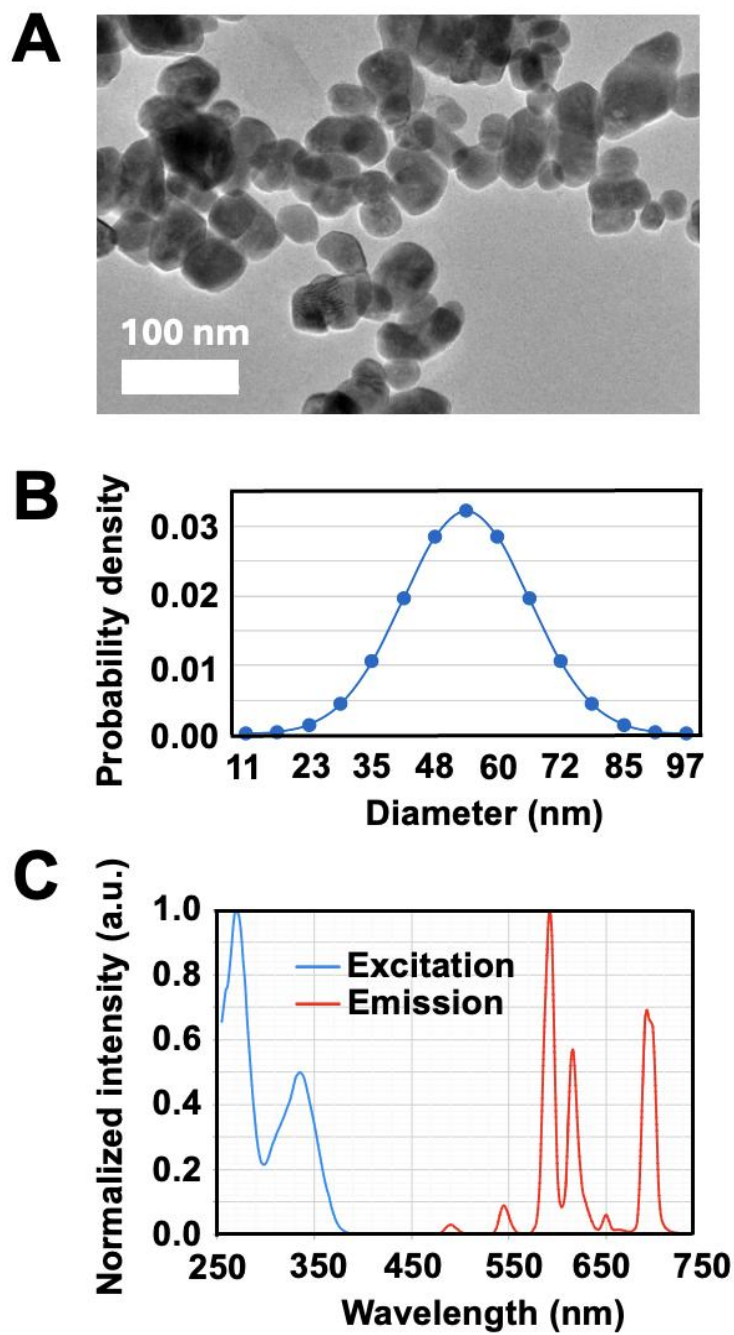

**Figure 1.** Characterization of the Bright-Dtech™ 614 - EuNPs showing **(A)** TEM image of the nanoparticles, **(B)** size distribution (n=100), and **(C)** excitation and emission spectrum.

The luminescence quantum yield of the Bright-Dtech™ 614 EuNPs was quantified at 70%, a commendable value compared to other Eu-based nanoparticles [25]. This high quantum yield ensures bright signal emission at the TL of the LFIA strip, making it detectable by the naked eye with the appropriate excitation source. The photophysical properties of these nanoparticles are well-suited for imaging with the western blot module of the Molecular Devices SpectraMax ID5 plate reader. Given the excellent compatibility of Bright-Dtech™ 614 EuNPs with LFIA, we proceeded to conjugate them with anti-LDH antibodies via direct adsorption. The conjugation efficacy was determined to be 90%, as quantified by measuring the residual antibodies in the supernatant. This conjugation efficiency ensures an optimal number of available detection bioreceptors, thereby maximizing assay sensitivity.

*3.2. Optimization of the EuNPs-LFIA for h-LDH detection*

Following the characterization of Bright-Dtech™ 614 EuNPs, the next step was to utilize them as signal reporters in the LFIA. This process involved conjugating the EuNPs with anti-LDH antibodies at optimal ratios to maximize assay sensitivity. The conjugation was performed at particle ratios of 1:50, 1:100, and 1:150, and the resulting complexes were tested on strips functionalized with capture anti-LDH antibodies at concentrations of 0.5, 0.8, and 1.0 mg/mL in the TL. Serial dilutions of h-LDH (0, 0.1, 20, and 40 ng/mL) were analyzed using these strips, and the normalized signal intensity at the TL was recorded. As expected, signal intensity increased with higher concentrations of h-LDH; however, it was observed that higher particle ratios of antibodies led to decreased signal intensity, probably due to steric hindrance (Figure 2A). Signals became more apparent to the naked eye with strips containing higher concentrations of capture antibodies in the TL; nevertheless, this also resulted in increased non-specific signal generation when analyzing blank samples (Figure S2).

A particle ratio of 1:100 with 0.5 mg/mL of capture anti-LDH antibody exhibited the lowest non-specific signal among the tested conditions. A particle ratio 1:150 with 0.8 mg/mL of capture anti-LDH antibody produced the most significant signal difference between the 0 and 0.1 ng/mL h-LDH samples. These two conditions were selected for further optimization of the LFIA.

In the second optimization step, we evaluated the impact of EuNP concentration on assay sensitivity. EuNPs at concentrations of 2.4, 4.8, and $9.2 \times 10^{-11}$ M were conjugated to antibodies at ratios of 1:100 and 1:150 and tested on LFIA strips with capture anti-LDH antibody concentrations of 0.5 and 0.8 mg/mL. Low concentrations of h-LDH (0, 0.05, and 0.1 ng/mL) were analyzed to determine which conditions provided the highest assay sensitivity. No significant differences in TL signal were observed when testing increasing concentrations of EuNPs with strips functionalized with 0.5 mg/mL of capture anti-LDH antibody (Figure 2B); thus, this condition was excluded from further consideration.

Conversely, when using strips with 0.8 mg/mL of capture anti-LDH antibody, the application of $2.4 \times 10^{-11}$ M of EuNPs generated an increase in TL signal, even at a detection threshold of 0.05 ng/mL of h-LDH, significantly distinguishing it from the blank sample. Consequently, we selected the optimal conditions of $4.8 \times 10^{-11}$ M of EuNPs, a 1:150 EuNP-antibody ratio, and a capture antibody concentration of 0.8 mg/mL for further assays.

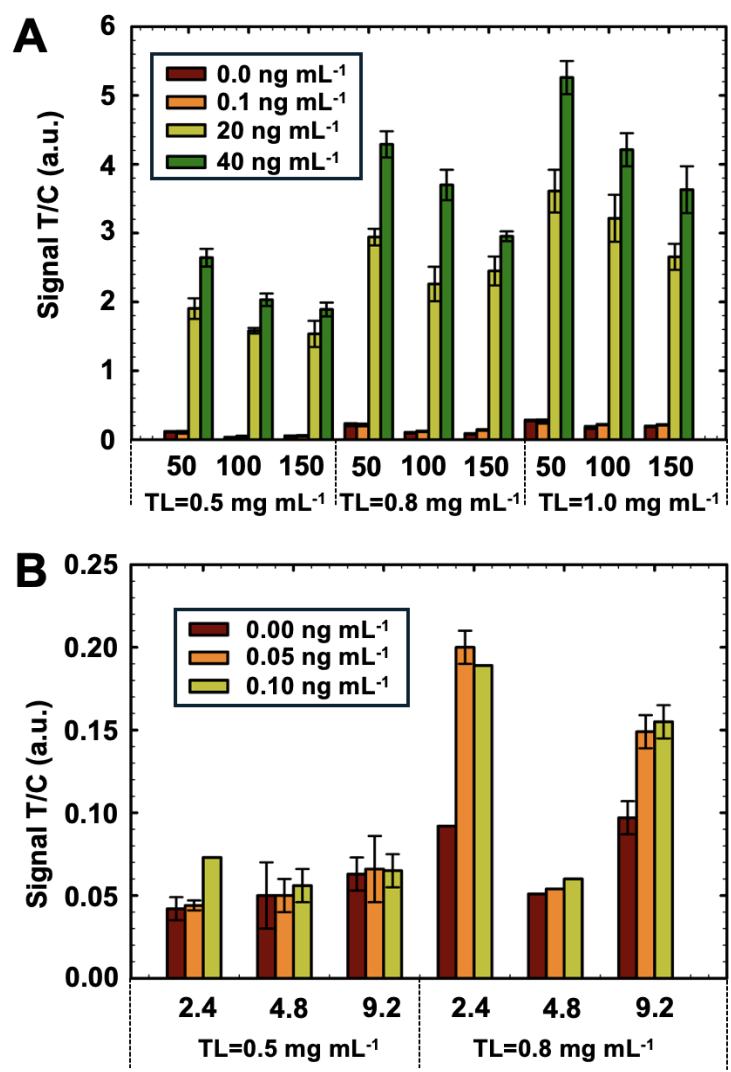

**Figure 2.** Optimization of the LFIA. **(A)** Evaluation of 50, 100, and 150 anti-LDH antibody particles per EuNP and 0.5, 0.8, and 1.0 mg mL$^{-1}$ of capture anti-LDH antibody in the TL for detecting 0, 0.1, 20, and 40 ng mL$^{-1}$ of h-LDH in working buffer (n=3). **(B)** Evaluation of 2.4, 4.8, and 9.2E-11 M of EuNPs and 0.5 and 0.8 mg mL$^{-1}$ of capture anti-LDH antibody in the TL for detecting 0, 0.05, and 0.01 ng mL$^{-1}$ of h-LDH in working buffer (n=3).

*3.3. Calibration curve for h-LDH detection*

Once the assay was optimized, we determined the analytical parameters of the developed test by performing a calibration curve using serial dilutions of h-LDH spiked in the working buffer. Figure 3A presents images of the LFIA strips recorded with the reader after completing the calibration curve. The signal in the TL exhibited a dose-response relationship, increasing steadily with higher concentrations of h-LDH. Signal saturation occurred at 40 ng/mL of h-LDH, while a 100 ng/mL concentration resulted in a hook effect (data not shown). A visual limit of detection (vLOD) was established at 0.4 ng/mL of h-LDH, as a panel of four members could distinguish the signal in strip 4 (interrogated with 0.4 ng/mL) from the signal in strip 1 (blank sample). Moreover, the signal in the CL remained consistently high, independent of the h-LDH concentration, thus serving as a reliable positive control for both qualitative and quantitative evaluations of the strips, aiding in TL signal normalization.

Figure 3B illustrates the plot generated from quantifying the signals in the TL using the reader. The quantified signal demonstrated a clear dose-response relationship across the 0.06 to 40 ng/mL (see the inset in Figure 3B). The collected data were fitted to a four-parameter logistic (sigmoidal) equation ($Y = 12.78 \times (\frac{5.55-X}{X-0.16})^{(\frac{-1}{1.31})}$), allowing for the determination of key analytical parameters such as the limit of detection (LoD), limit of quantification (LoQ), EC50, dynamic linear range (LDR), and the linear regression coefficient (r²), as summarized in Table S1. The LoD was determined to be 38 pg/mL of h-LDH, 686-fold and 15-fold lower than the previously reported limits for LFIA based on AuNPs and CNPs, respectively (Table 1). The assay demonstrated high reproducibility, with a mean intra-batch and inter-batch variability of 9.7 and 4.9 %, respectively, when analyzing h-LDH concentrations of 0, 1, and 10 ng mL$^{-1}$ (Table S2).

Given the reported threshold of h-LDH activity (233 U/L in serum, 602 mU/mg in normal breast tissue, and 630 mU/mg in tumor breast tissue), the sensitivity of our LFIA enables early breast cancer diagnosis and precise monitoring of therapies targeting h-LDH. With a detection limit of 38 pg/mL, our assay is 1,000,000 times more sensitive than the serum threshold (233 U/L), allowing detection and monitoring of h-LDH levels across various biological samples.

The ability to reach a sensitivity of 271 fM (considering h-LDH is 140 kDa) in LFIA by using fluorescent Bright-Dtech™ EuNPs, significantly enhances the applicability of this rapid diagnostic test for detecting disease biomarkers at ultra-low concentrations in serum. Compared to other LFAs utilizing Europium-chelated nano/microparticles as time-resolved fluorescence reporters, only one approach achieves significantly lower sensitivity. However, this enhanced sensitivity is attributed to the incorporation of PCR amplification to increase the target analyte concentration before the LFIA application (Table S3).

This assay sensitivity is particularly advantageous compared to time-consuming methods like well-plate ELISA, which cannot be performed at the point of care. Notably, the LoD achieved was 2.9-fold lower than that obtained using the gold standard method (well-plate ELISA) (Table 1), which typically requires a two-hour procedure. Furthermore, the developed assay demonstrated reproducibility, with an average relative standard deviation (RSD) of 6 ± 3% in the signals generated among replicates, which is critical for accurately quantifying such low levels of h-LDH.

**Table 1. Comparison of the h-LDH detection methods**

| Detection method | Assay time (min) | Applicability at the point of care | Detection limit (ng mL$^{-1}$) | Ref. |
|---|---|---|---|---|
| Enzyme-linked immunosorbent assay | 120 | No | 0.11 | [17] |
| Colorimetric immunosensor | 40 | No | 9.2 | [26] |
| Disk-based one-dimensional photonic crystal slabs | >30 | No | 1.8 | [27] |
| Fluorescence quenching | 20 | No | 2500 | [28] |
| CNPs-based Lateral Flow Immunoassay | 10 | Yes | 0.57 | [17] |
| EuNPs-based Lateral Flow Immunoassay | 20 | Yes | 0.038 | This work |

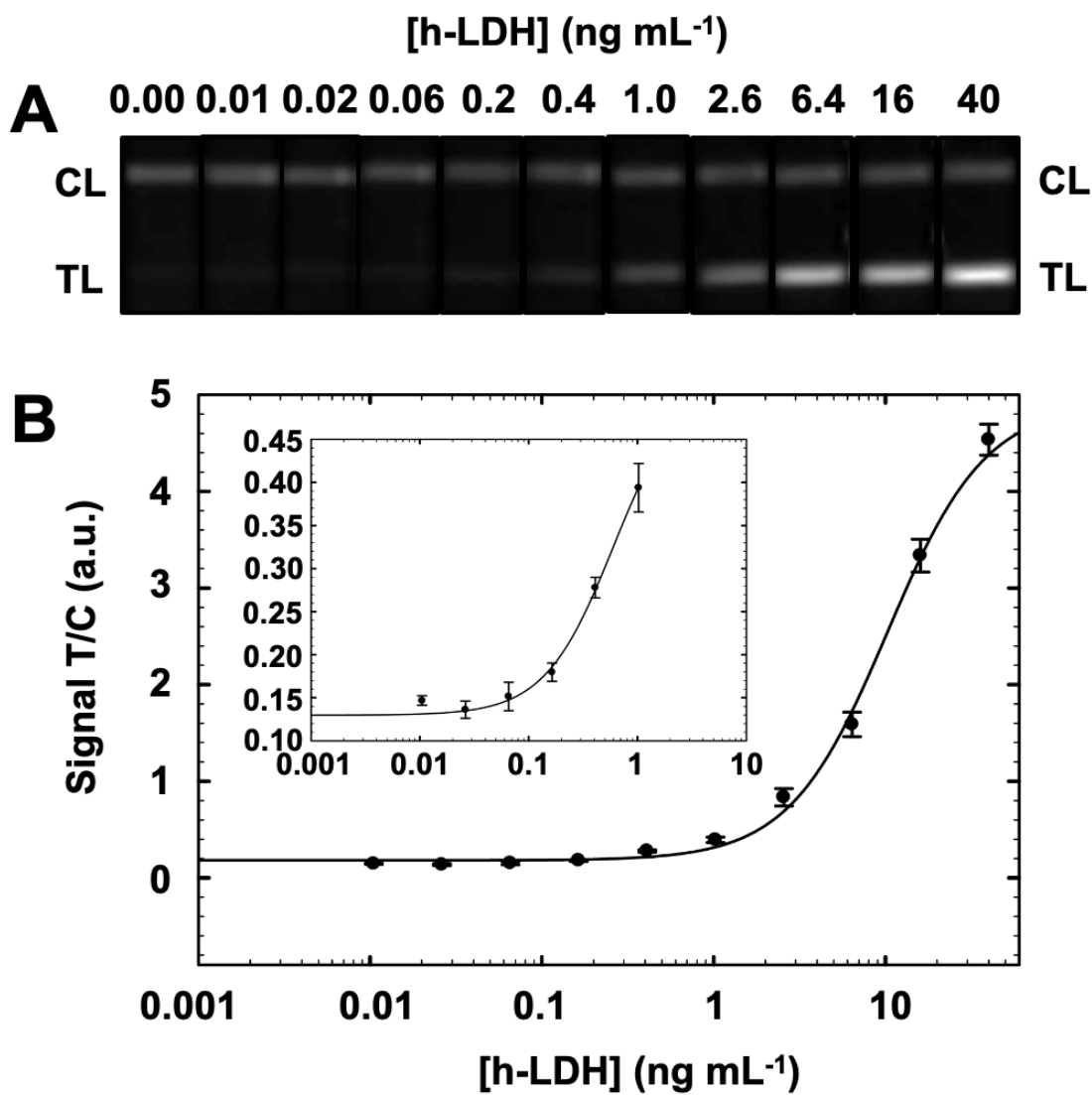

**Figure 3.** (**A**) Images of the EuNPs-based LFIA strips after detecting serial dilutions of h-LDH in working buffer 1) 0.00, 2) 0.01, 3) 0.02, 4) 0.06, 5) 0.16, 6) 0.40, 7) 1.02, 8) 2.56, 9) 6.4, 10) 16.00, 11) 40.00 ng mL$^{-1}$). (**B**) Calibration curve plot representing the normalized signal in the TL achieved when detecting serial dilutions of h-LDH standard in working buffer with EuNPs-based LFIA (n = 4). Inset showing the calibration curve plot between 0.001 and 10 ng mL$^{-1}$ of h-LDH.

### 3.4. Recovery studies with serum samples

To evaluate the applicability of the developed assay for complex samples, h-LDH was spiked at various concentrations in serum and analyzed using LFIA strips. Each sample was tested in quadruplicate to ensure accuracy, precision, and reproducibility. An 8-fold dilution of the serum samples provided optimal migration of the fluorescent nanoparticles along the strip, resulting in a high signal in both the test and control lines. The signals were quantified, and the h-LDH concentration was determined using the calibration curve equation. Importantly, no matrix effects were observed in the strips when analyzing serum samples with no h-LDH and levels near the IC50 ± 50% of the calibration curve (Figure 3B). The assay achieved an average recovery of 108 ± 11% (Table 2), confirming the developed LFIA can be used to accurately quantify the levels of h-LDH in complex samples, such as serum

Furthermore, the analysis of these complex samples did not disrupt the signal generated in the strips, which remained consistent across replicates, demonstrating an average RSD of 7.34%. These results revealed that Bright-Dtech™ 614 Eu nanoparticles can be effectively utilized in clinical settings to analyze serum samples from patients using LFIA.

**Table 2. Recovery results when analyzing h-LDH spiked serum samples with EuNPs-based LFIA.**

| Sample | Expected [h-LDH] (ng mL$^{-1}$) | Detected [h-LDH] (ng mL$^{-1}$) | Recovery (%) |
|---|---|---|---|
| 1 | 0 | < LoD | - |
| 2 | 0.6 | 0.8 ± 0.1 | 121 |
| 3 | 1.2 | 1.2 ± 0.1 | 100 |
| 4 | 2.5 | 2.4 ± 0.2 | 94 |
| 5 | 5.0 | 5.3 ± 0.6 | 105 |
| 6 | 6.2 | 7.1 ± 0.2 | 113 |
| 7 | 10.0 | 11.7 ± 0.7 | 117 |
| 8 | 12.5 | 14.3 ± 1.0 | 114 |
| 9 | 20.0 | 23.2 ± 2.6 | 116 |
| 10 | 25.0 | 21.9 ± 4.8 | 88 |
| **Mean** | | | **108 ± 11** |

**4. Conclusions**

This study marks a significant advancement in lateral flow immunoassays by successfully using Bright-Dtech™ 614 Eu nanoparticles to enhance the sensitivity for detecting the cancer biomarker h-LDH. Our findings demonstrate that Eu nanoparticles, with their remarkable 70% quantum yield and 90% conjugation efficacy with anti-LDH antibodies, can serve as highly effective reporters for LFIA. By optimizing the nanoparticle-to-antibody ratio, EuNP, and capture antibody concentrations, we achieved a detection limit of 38 pg mL$^{-1}$, a remarkable sensitivity enhancement by factors of 686, 15, and 2.9 compared to traditional LFIA platforms using AuNPs, CNPs, and even the well-established ELISA, respectively. Notably, the LFIA developed in this study reaches the femtomolar sensitivity range, underscoring its potential for detecting trace levels of biomarkers and advancing early-stage cancer diagnosis. While a bench-top plate reader was used to quantify signals from the strips, the developed assay is also suitable for point-of-care applications using commercially available portable LFA readers, such as the IPeak®Eu (IUL). With up to 4 hours of autonomy, this battery-powered device is specifically designed to analyze Europium-based LFAs using 365/615 nm excitation/emission. Furthermore, the assay demonstrated excellent reliability in serum samples, exhibiting high recovery accuracy and reproducibility, reinforcing its potential for clinical applications. With a laboratory-scale fabrication cost of €0.22 per strip (22% lower than AuNPs-based LFIAs) (Table S4), this work marks a significant step forward in advancing LFIA technology for precision medicine. This advancement facilitates the development of more sensitive, rapid, and accessible diagnostics, offering significant potential to improve patient outcomes and revolutionize disease management.

**CrediT authorship contribution statement**

**J.L.**: Project Administration, Investigation, Methodology, Formal Analysis, Validation. **Y.D.B.O**: Investigation, Methodology, Formal Analysis, Writing - original draft. **A.S.T**: Project Administration, Data curation, Formal analysis, Validation, Visualization, Writing - original draft. **L.C.:** Investigation, Methodology, Formal analysis, Writing - review & editing. **M.S.**: Project Administration, Investigation, Methodology, Formal analysis, Validation. **J.G.**: Project Administration, Conceptualization, Writing - review & editing. **A.M**: Funding acquisition, Writing - review & editing. **S.M**: Conceptualization, Funding acquisition, Writing - review & editing.

**Acknowledgments**

Y.D.B.O. acknowledges the financial support of the Ministerio de Ciencia Tecnología e Innovación Scholarship Program No. 8852 of Colombia. A.S. acknowledges Generalitat Valenciana and the European Social Fund for the CIAPOS fellowship (CIAPOS/2022/7).

**Declaration of interests**

The authors declare that they have no known competing financial interests or personal relationships that could have appeared to influence the work reported in this paper. Three authors (Juliette Lajoux, Mohamadou Sy, and Joan Goetz) are employees of Poly-Dtech corporation. Poly-Dtech holds a patent on Bright-Dtech™ nanoparticles. Poly-Dtech commercializes the Link-Dtech™ 614 – Eu commercial kit.

**References**

[1]  C. Parolo, A. Sena-Torralba, J.F. Bergua, E. Calucho, C. Fuentes-Chust, L. Hu, L. Rivas, R. Álvarez-Diduk, E.P. Nguyen, S. Cinti, D. Quesada-González, A. Merkoçi, Tutorial: design and fabrication of nanoparticle-based lateral-flow immunoassays, Nat Protoc 15 (2020) 3788–3816. https://doi.org/10.1038/s41596-020-0357-x.

[2]  D. Quesada-González, A. Merkoçi, Nanoparticle-based lateral flow biosensors, Biosensors and Bioelectronics 73 (2015) 47–63. https://doi.org/10.1016/j.bios.2015.05.050.

[3]  F. Di Nardo, M. Chiarello, S. Cavalera, C. Baggiani, L. Anfossi, Ten Years of Lateral Flow Immunoassay Technique Applications: Trends, Challenges and Future Perspectives, Sensors (Basel) 21 (2021) 5185. https://doi.org/10.3390/s21155185.

[4]  B.D. Grant, C.E. Anderson, J.R. Williford, L.F. Alonzo, V.A. Glukhova, D.S. Boyle, B.H. Weigl, K.P. Nichols, SARS-CoV-2 Coronavirus Nucleocapsid Antigen-Detecting Half-Strip Lateral Flow Assay Toward the Development of Point of Care Tests Using Commercially Available Reagents, Anal Chem 92 (2020) 11305–11309. https://doi.org/10.1021/acs.analchem.0c01975.

[5]  P.E. Geyer, L.M. Holdt, D. Teupser, M. Mann, Revisiting biomarker discovery by plasma proteomics, Mol Syst Biol 13 (2017) 942. https://doi.org/10.15252/msb.20156297.

[6]  A. Scohy, A. Anantharajah, M. Bodéus, B. Kabamba-Mukadi, A. Verroken, H. Rodriguez-Villalobos, Low performance of rapid antigen detection test as frontline testing for COVID-19 diagnosis, J Clin Virol 129 (2020) 104455. https://doi.org/10.1016/j.jcv.2020.104455.

[7]  A. Sena-Torralba, R. Álvarez-Diduk, C. Parolo, A. Piper, A. Merkoçi, Toward Next Generation Lateral Flow Assays: Integration of Nanomaterials, Chem Rev 122 (2022) 14881–14910. https://doi.org/10.1021/acs.chemrev.1c01012.

[8]  Z. Gao, H. Ye, D. Tang, J. Tao, S. Habibi, A. Minerick, D. Tang, X. Xia, Platinum-Decorated Gold Nanoparticles with Dual Functionalities for Ultrasensitive Colorimetric in Vitro Diagnostics, Nano Lett 17 (2017) 5572–5579. https://doi.org/10.1021/acs.nanolett.7b02385.

[9]  X. Song, M. Knotts, Time-resolved luminescent lateral flow assay technology, Analytica Chimica Acta 626 (2008) 186–192. https://doi.org/10.1016/j.aca.2008.08.006.

[10] K.Y. Zhang, Q. Yu, H. Wei, S. Liu, Q. Zhao, W. Huang, Long-Lived Emissive Probes for Time-Resolved Photoluminescence Bioimaging and Biosensing, Chem. Rev. 118 (2018) 1770–1839. https://doi.org/10.1021/acs.chemrev.7b00425.

[11] J. Yuan, G. Wang, Lanthanide Complex-Based Fluorescence Label for Time-Resolved Fluorescence Bioassay, J Fluoresc 15 (2005) 559–568. https://doi.org/10.1007/s10895-005-2829-3.

[12] E. Soini, T. Lövgren, C.B. Reimer, Time-Resolved Fluorescence of Lanthanide Probes and Applications in Biotechnology, C R C Critical Reviews in Analytical Chemistry 18 (1987) 105–154. https://doi.org/10.1080/10408348708542802.

[13] Q. Liu, S. Cheng, R. Chen, J. Ke, Y. Liu, Y. Li, W. Feng, F. Li, Near-Infrared Lanthanide-Doped Nanoparticles for a Low Interference Lateral Flow Immunoassay Test, ACS Appl. Mater. Interfaces 12 (2020) 4358–4365. https://doi.org/10.1021/acsami.9b22449.

[14] T. Salminen, E. Juntunen, S.M. Talha, K. Pettersson, High-sensitivity lateral flow immunoassay with a fluorescent lanthanide nanoparticle label, Journal of Immunological Methods 465 (2019) 39–44. https://doi.org/10.1016/j.jim.2018.12.001.

[15] Z. Chen, Z. Zhang, X. Zhai, Y. Li, L. Lin, H. Zhao, L. Bian, P. Li, L. Yu, Y. Wu, G. Lin, Rapid and Sensitive Detection of anti-SARS-CoV-2 IgG, Using Lanthanide-Doped Nanoparticles-Based Lateral Flow Immunoassay, Anal. Chem. 92 (2020) 7226–7231. https://doi.org/10.1021/acs.analchem.0c00784.

[16] Bright-Dtech$^{TM}$: Fluorescent dyes - Poly-Dtech, (n.d.). https://poly-dtech.com/products/nanoparticles/bright-dtech/ (accessed October 17, 2024).

[17] G. Feugray, C. Dumesnil, M. Grall, Y. Benhamou, H. Girot, J. Fettig, V. Brunel, P. Billoir, Lactate dehydrogenase and hemolysis index to predict vaso-occlusive crisis in sickle cell disease, Sci Rep 13 (2023) 21198. https://doi.org/10.1038/s41598-023-48324-w.

[18] G.J. Kato, V. McGowan, R.F. Machado, J.A. Little, J. Taylor, C.R. Morris, J.S. Nichols, X. Wang, M. Poljakovic, S.M. Morris, M.T. Gladwin, Lactate dehydrogenase as a biomarker of hemolysis-associated nitric oxide resistance, priapism, leg ulceration, pulmonary hypertension, and death in patients with sickle cell disease, Blood 107 (2006) 2279–2285. https://doi.org/10.1182/blood-2005-06-2373.

[19] A. Comandatore, M. Franczak, R.T. Smolenski, L. Morelli, G.J. Peters, E. Giovannetti, Lactate Dehydrogenase and its clinical significance in pancreatic and thoracic cancers, Semin Cancer Biol 86 (2022) 93–100. https://doi.org/10.1016/j.semcancer.2022.09.001.

[20] K. Augoff, A. Hryniewicz-Jankowska, R. Tabola, Lactate dehydrogenase 5: an old friend and a new hope in the war on cancer, Cancer Lett 358 (2015) 1–7. https://doi.org/10.1016/j.canlet.2014.12.035.

[21] Y. Rong, W. Wu, X. Ni, T. Kuang, D. Jin, D. Wang, W. Lou, Lactate dehydrogenase A is overexpressed in pancreatic cancer and promotes the growth of pancreatic cancer cells, Tumour Biol 34 (2013) 1523–1530. https://doi.org/10.1007/s13277-013-0679-1.

[22] Y.D. Banguera-Ordoñez, A. Sena-Torralba, P. Quintero-Campos, Á. Maquieira, S. Morais, Smartphone-based lateral flow immunoassay for sensitive determination

of lactate dehydrogenase at the point of care, Talanta 281 (2024) 126803. https://doi.org/10.1016/j.talanta.2024.126803.

[23] D.A. Armbruster, T. Pry, Limit of blank, limit of detection and limit of quantitation, Clin Biochem Rev 29 Suppl 1 (2008) S49-52.

[24] F. Ricci, A. Vallée-Bélisle, K.W. Plaxco, High-precision, in vitro validation of the sequestration mechanism for generating ultrasensitive dose-response curves in regulatory networks, PLoS Comput Biol 7 (2011) e1002171. https://doi.org/10.1371/journal.pcbi.1002171.

[25] D. Tu, W. Zheng, P. Huang, X. Chen, Europium-activated luminescent nanoprobes: From fundamentals to bioapplications, Coordination Chemistry Reviews 378 (2019) 104–120. https://doi.org/10.1016/j.ccr.2017.10.027.

[26] I.F. Pinto, R.R.G. Soares, M.E.-L. Mäkinen, V. Chotteau, A. Russom, Multiplexed Microfluidic Cartridge for At-Line Protein Monitoring in Mammalian Cell Culture Processes for Biopharmaceutical Production, ACS Sens 6 (2021) 842–851. https://doi.org/10.1021/acssensors.0c01884.

[27] G. Sancho-Fornes, M. Avella-Oliver, J. Carrascosa, E. Fernandez, E.M. Brun, Á. Maquieira, Disk-based one-dimensional photonic crystal slabs for label-free immunosensing, Biosens Bioelectron 126 (2019) 315–323. https://doi.org/10.1016/j.bios.2018.11.005.

[28] X. Ren, L. Yang, F. Tang, C. Yan, J. Ren, Enzyme biosensor based on NAD-sensitive quantum dots, Biosens Bioelectron 26 (2010) 271–274. https://doi.org/10.1016/j.bios.2010.05.014.